\documentclass[aps,prb,twocolumn,showpacs,amsmath,superscriptaddress]{revtex4}
\usepackage{graphicx}
\usepackage{bm}

\begin{document}

\title{Longitudinal Proximity Effects in Superconducting Transition-Edge Sensors}

\author{John E.\ Sadleir}
\email[]{john.e.sadleir@nasa.gov}
\affiliation{Department of Physics, University of Illinois, 1110 West Green Street, Urbana, IL 61801-3080}
\affiliation{NASA Goddard Space Flight Center, 8800 Greenbelt Road, Greenbelt, MD 207701}

\author{Stephen J.\ Smith}
\affiliation{%
   NASA Goddard Space Flight Center, 8800 Greenbelt Road, Greenbelt, MD 207701}
\author{Simon R. Bandler}
\affiliation{%
   NASA Goddard Space Flight Center, 8800 Greenbelt Road, Greenbelt, MD 207701}
\author{James A. Chervenak}
\affiliation{%
   NASA Goddard Space Flight Center, 8800 Greenbelt Road, Greenbelt, MD 207701}

\author{John R.\ Clem}
\affiliation{%
   Ames Laboratory and Department of Physics and Astronomy, Iowa State University, Ames, Iowa, 50011--3160}

\date{\today}

\begin{abstract} 
We have found experimentally that the critical current of a square superconducting transition-edge sensor (TES) depends exponentially upon the side length $L$ and the square root of the temperature $T$.  As a consequence, the effective transition temperature $T_c$ of the TES is current-dependent and at fixed current scales as $1/L^2$.  We also have found that the critical current can show clear Fraunhofer-like oscillations in an applied  magnetic field, similar to those found in Josephson junctions. The observed behavior has a natural theoretical explanation in terms of longitudinal proximity effects if the TES is regarded as a weak link between superconducting leads.  We have  observed  the proximity effect in these devices over extraordinarily long lengths exceeding 100 $\mu$m.
\end{abstract}

\pacs{74.25.-q,74.78.Bz,74.25.Op}
\maketitle

A superconductor cooled through its transition temperature $T_c$ while carrying a finite dc bias current undergoes an abrupt decrease in electrical resistance from its normal-state value $R_N$ to zero.  Superconducting transition-edge sensors (TESs) exploit this sharp transition; these devices are highly sensitive resistive thermometers used for precise thermal energy measurements.\cite{IrwinHilton05}  TES microcalorimeters have been developed with measured energy resolutions in the X-ray and gamma-ray band of 
$\Delta E=1.8\pm$0.2 eV FWHM at 6 keV, \cite{Bandler08} 
and 
$\Delta E=22$ eV FWHM at 97 keV, \cite{Bacrania09}
respectively--- with the latter result at present the largest reported $E/\Delta E$ of any non-dispersive photon spectrometer.  TESs are successfully used across much of the electromagnetic spectrum, measuring the energy of single-photon absorption events from infrared to gamma-ray energies and photon fluxes out to the microwave range.\cite{IrwinHilton05}  Despite these experimental successes, the dominant physics governing TESs biased in the superconducting phase transition remains poorly understood.\cite{IrwinHilton05}

To achieve high energy resolution it is important to control both the TES's $T_c$ and its transition width $\Delta T_c$.  Because the energy resolution of calorimeters improves with decreasing temperature, they are typically designed to operate at temperatures around 0.1 K.  For a TES, this requires a superconductor with $T_c$ in that range.  While there exist a few suitable elemental superconductors, the best results have been achieved using proximity-coupled, superconductor/normal-metal (S/N) bilayers\cite{Bandler08,Bacrania09}, for which $T_c$ is tuned by selection of the thicknesses of the S and N layers. \cite{Martinis00}



There have been a variety of models\cite{Fraser04,Seidel04,Martinis00,Luukanen03,LindemanPerc06}  used to explain the noise, $T_c$, and $\Delta T_c$ in TES bilayers, all assuming spatially uniform devices.  Though some have been shown to be consistent with certain aspects of particular devices, they do not explain measured $T_c$ and $\Delta T_c$ in S/N bilayer TESs generally.  

In this paper we emphasize the importance of a phenomenon that so far has been neglected in previous theoretical studies of TESs:  the longitudinal proximity effect.  Since the square bilayers at the heart of the TES are connected at opposite ends to superconducting leads with transition temperatures  well above the intrinsic transition temperature of the bilayers, superconductivity is induced longitudinally into the bilayers via the proximity effect.  As we shall explain later, many of the basic properties of our TES structures are well described by regarding them as SS$'$S or SN$'$S weak links.\cite{SS'S,Clarke69,Likharev79} 

In this paper we report the properties of TESs based on square ($L \times L$) electron-beam-deposited Mo/Au bilayers consisting of 55 nm Mo layers ($T_c \sim $ 0.9 K) to which 210 nm of Au is added.  The square side lengths $L$  range from 8 $\mu$m to 290 $\mu$m, and the normal-state resistance per square is $R_N=17.2\pm$0.5 m$\Omega$.   The  bilayers are connected at opposite ends to  Mo/Nb leads having measured superconducting transition temperatures of 3.5 and 7.1 K.\cite{LeadsTcComment}   Further details on the device fabrication process can be found in Ref.\ \onlinecite{Chervenak04}.

Our measurements are made in an adiabaticÊ demagnetization refrigerator (ADR) with mu-metal and Nb enclosures providingÊ magnetic shielding for the TES devices and SQUID electronics.  The magnetic field normal to the TES device plane is controlled by a superconducting coil with the field value determined from the coil geometry and current.  Measurements of the TES resistance $R$ are made byÊ applying a sinusoidal current of frequency 5-10 Hz and amplitude $I_{bias} \sim$ 50-250 nA, with zero dc component, to the TES in parallel with a 0.2 m$\Omega$ shunt resistor ($R_{sh}$).  The time-dependent TES current is measured with a SQUID feedback circuit with input coil in series with TES. When $I_{bias}$ is less than the TES critical current $I_c$,  $R$ is zero, and all the ac current flows through the TES.  However, when $I_{bias}>I_c$ and $R > 0$ during part of the ac cycle, the TES current becomes non-sinusoidal, and its maximum value $I$ becomes less than $I_{bias}$.  The TES resistance $R$ at the TES current $I$ is then determined from $R=R_{sh} (I_{bias}-I)/I$.

The critical current $I_c$ is measured, with the ADR held at constant temperature, by ramping the dc bias current from zero and defining $I_c$ as the TES current at the first measured finite resistance ($R \sim 10\; \mu\Omega$) across the TES.  Record averaging is used at higher temperatures where $I_c$ becomes small. 

\begin{figure}
\includegraphics[width=8cm]{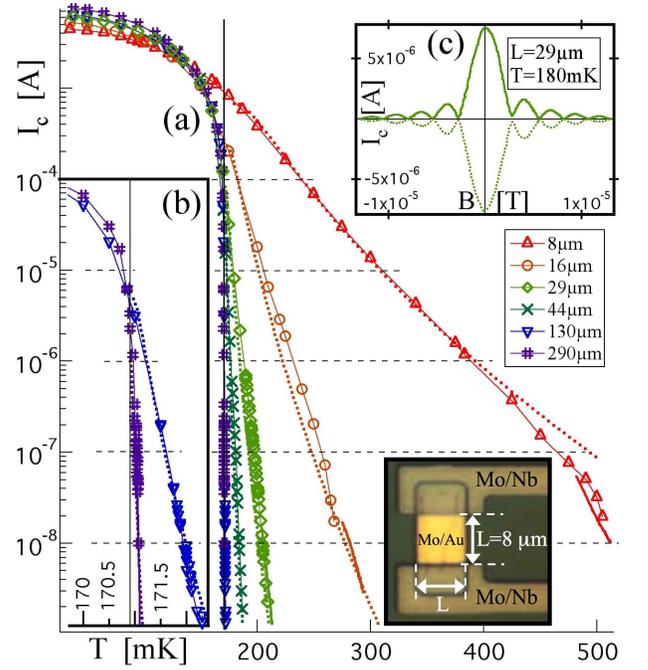}
\caption{%
(Color online)  (a) and inset (b) Measured (solid lines and markers) and theoretical (dotted curves) critical current $I_c$  versus temperature $T$ for square TESs with side lengths $L$ ranging from 8 to 290 $\mu$m.  The bold continuous segments at the lowest currents are obtained by record averaging.  The intrinsic transition temperature of the Mo/Au bilayer weak links is $T_{cw} = 170.9 \pm$0.1 mK (thin vertical lines).  For $T$ somewhat larger than $T_{cw}$, $I_c$ decays approximately exponentially with the square root of $T-T_{cw}$.  $I_c$ also depends strongly upon $L$, which is particularly noticeable for the smaller devices.   $T$ and $L$ values of the constant current contours (horizontal dashed lines) are plotted in Fig.\ 2(a).  Inset (c) shows $I_c$ vs applied field for the $L=29$ $\mu$m device showing Fraunhofer-like oscillations, similar to those seen in Josephson junctions, providing further evidence that the TES exhibits weak-link behavior.}
\label{Icexpt}
\end{figure}

The solid curves in Fig.\ \ref{Icexpt}(a) and (b) show measurements of the critical current $I_c$ over seven decades vs temperature $T$.  Note that although we find the intrinsic transition temperature of the Mo/Au bilayer is $T_{cw} = 170.9\pm$0.1 mK,  at very low currents a zero-resistance state is measured up to much higher temperatures as the TES size is reduced, three times $T_{cw}$ for $L$ = 8 $\mu$m.  On the other hand, for the larger TES sizes ($L$=130 and 290 $\mu$m) the critical current $I_c(T)$ decreases rapidly with $T$ near $T_{cw}$.  The observed $I_c$ behavior as functions of both $T$ and the length $L$ provides strong evidence that our TESs behave as weak-link devices.  The dotted curves in (a) and (b) show calculated values of $I_c$ using the Ginzburg-Landau theory described below.  In addition, at appropriately chosen temperatures, the critical currents of these devices exhibit Fraunhofer-like oscillations as a function of an applied magnetic field, behavior characteristic of Josephson weak links.\cite{Clarke69,Barone82,Miller84,Dobrosavljevic93}  See Fig.\ \ref{Icexpt}(c) for an example.

\begin{figure}
\includegraphics[width=8cm]{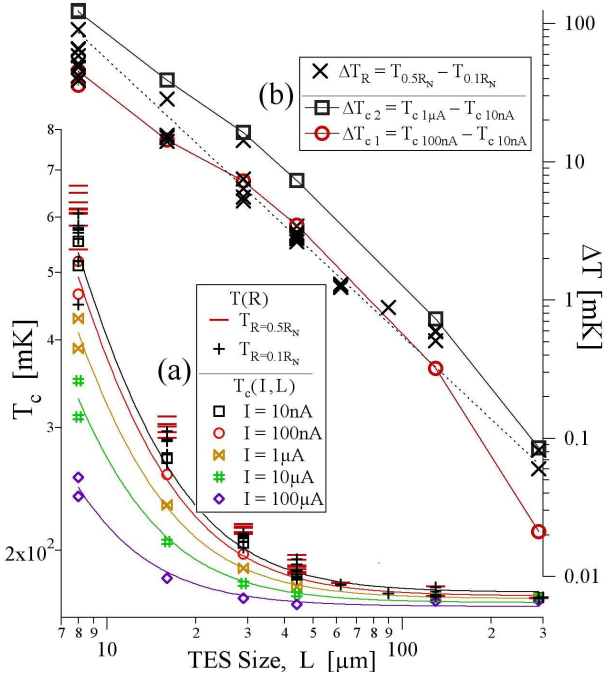}
\caption{%
(a) Measurements of the effective transition temperature $T_c$ at different currents and lengths.  Markers $T_c(I,L)$ give the effective $T_c$ from constant current contours of the $I_c(T,L)$ data in Fig.\ 1, with solid curves being $1/L^2$ fits for each current level.  Markers $T(R)$ give temperatures where  $R = 0.1\,R_N$ and $0.5\,R_N$.  (b) Three different measures of the transition width defined as differences between pairs of corresponding points shown in (a), as labeled $\Delta T_R,$ $\Delta T_{c\: 1}$, and $\Delta T_{c\: 2}$, showing $1/L^2$ scaling (dotted line).}
\label{TcandDeltaTc}
\end{figure}

Because $I_c$ depends upon $T$ and $L$, the effective transition temperature $T_c$ of the TES (the temperature at which an electrical resistance first appears, i.e., $R \sim 10\; \mu\Omega$) is both current-dependent and length-dependent.  Figure \ref{TcandDeltaTc}(a) exhibits these effects.  The points labeled $T_c(I,L)$ are the effective transition temperatures at five different current levels (10 nA to 100 $\mu$A) for the data in Fig.\ \ref{Icexpt}(a) and (b), showing that $T_c - T_{cw}$ for each current level  scales approximately as $1/L^2$ (solid curve fits) for $L$ ranging from 8 to 290 $\mu$m.  For each $L$, $T_c - T_{cw}$ depends upon the current.  

Also shown in Fig.\ 2(a) are  temperatures $T_{R=0.1\,R_N}$ and  $T_{R=0.5\,R_N}$ for which the resistances are $R= 0.1\,R_N$ and $0.5\,R_N$, respectively, from which we define $\Delta T_R=T_{0.5R_N}-T_{0.1R_N}$.  We also define transition widths from the $I_c$ measurements $\Delta T_{c\: 1}=T_{c\:100nA} -T_{c\: 10nA}$ and $\Delta T_{c\: 2}=T_{c\: 1\mu A}-T_{c\:10nA}$.  In Fig.\ \ref{TcandDeltaTc}(b) we show that these three measures of the transition width all  vary approximately as $1/L^2$, shown by the dotted line.  It also follows that $T_c - T_{cw}$ scales linearly with the transition width.

Weak links in various SS$'$S or SN$'$S weak-link structures have been studied experimentally and theoretically by numerous authors.  However, here we use a simple version of Ginzburg-Landau (GL) theory\cite{deGennes,StJames69} to explain the results shown in Figs.\ \ref{Icexpt}
and \ref{TcandDeltaTc}.  This theory describes the complex order parameter $\psi(\bm r)$, whose absolute square $|\psi(\bm r)|^2$ is the superfluid density in the weak link.  We employ the substitution
$\psi = \psi_r f e^{i \gamma},$
where $\psi_r$ is the magnitude of the order parameter at the reference points $x = \pm L/2$ adjacent to the leads, $f=|\psi |/\psi_r$ is the normalized order parameter, and $\gamma$ is the phase.
At the reference points, the local value of $\psi_r^2$ is inversely proportional to the square of the local penetration depth $\lambda_r$ via
$\psi_r^2 = m/4 \mu_0 e^2\lambda_r^2,$ and a
characteristic reference current density $j_r$ can be defined via
$j_r=\phi_0/2 \pi \mu_0 \lambda_r^3$, where $\phi_0 = h/2e$.

Near the center of the weak link, where $f$ becomes very small, the local penetration depth $\lambda = \lambda_r/f$ becomes very large.  Moreover, in a thin film of thickness $d < \lambda$, magnetic fields and currents spread out over the two-dimensional screening length (or Pearl length\cite{Pearl64}) $\Lambda = 2 \lambda^2/d = 2 
\lambda_r^2/d f^2$.  
For all of our samples there is a range of temperatures $T$ far enough above $T_{cw}$  that at the center of the weak link $\Lambda \gg L$ and the current density is ${\bm j} = \hat x j_x = \hat x I/Wd$, where $I$ is the TES current.
The first and second GL equations given in Refs.\ \onlinecite{deGennes} and \onlinecite{StJames69} then depend only upon $x$ and can be written as 
\begin{equation}
-f'' +\frac{(t-1)}{\xi_w^2}f + \frac{\kappa^2}{\lambda_r^2}f^3 
+\frac{{\tilde j}^2}{\lambda_r^2 f^3}  = 0 
\label{GL1}
\end{equation}
and
\begin{equation}
\tilde j= j_x/j_r = -\lambda_r f^2 (\gamma'+2\pi A_x/\phi_0),
\label{GL2}
\end{equation}
where $t = T/T_{cw}$ is the reduced temperature, $\xi(T) = \xi_w/|t-1|^{1/2}$ is the temperature-dependent coherence length, $\kappa$ is the dimensionless Ginzburg-Landau parameter, $A_x$ is the vector potential, and the primes denote derivatives with respect to $x$.

In this paper we are concerned chiefly with weak-link behavior for which $f(x)$ is an even function of $x$ and has a minimum in the middle of the weak link, $f(0)=f_0,$ where $f'(0)=0$.  We can obtain an equation that determines how $f_0$ depends upon $L$, $t$, and $\tilde j$ by multiplying Eq.\ (\ref{GL1}) by $f'$, integrating the result,  multiplying by $f^2$, and taking the square root, which yields the following equation, valid for $0 \le x \le L/2$,
\begin{equation}
df^2/dx \!= \! P(f^2),
\label{df2bydx}
\end{equation}
where
\begin{equation}
P(f^2) \!\!=\! \! \sqrt{2(f^2 \!\!-\! \!f_0^2)\Big[\frac{\kappa^2}{\lambda_r^2}f^4 \!\!
+ \!\!\Big( \frac{2 (t \!- \!1)}{\xi_w^2} \!\!+\! \!\frac{\kappa^2 f_0^2}{\lambda_r^2}\Big)f^2 \!\!+ \!\!\frac{2\tilde j^2}{\lambda_r^2 f_0^2}\Big]},
\label{P}
\end{equation}
such that $f_0$ and $f(x)$ can be obtained from the integrals
\begin{equation}
\int_{f_0^2}^1\frac{df^2}{P(f^2)} = \frac{L}{2}\;\; \rm  and 
\label{f0integral}
\end{equation}
\begin{equation}
\int_{f_0^2}^{f(x)}\frac{df^2}{P(f^2)} = x.
\label{fintegral}
\end{equation}
The gauge-invariant phase difference across the weak link is\cite{Likharev75b}
\begin{equation}
\phi = -\int_{-L/2}^{L/2} \Big(\gamma'+\frac{2\pi A_x}{\phi_0}\Big)dx 
=\frac{2\tilde j}{\lambda_r}\int_{f_0^2}^1\frac{df^2}{f^2P(f^2)}.
\label{phiwintegral}
\end{equation}

The integrals in Eqs.\  (\ref{f0integral}), (\ref{fintegral}), and (\ref{phiwintegral}) can be evaluated numerically as in Ref.\ \onlinecite{Likharev75b} or in terms of elliptic integrals as in Refs.\ \onlinecite{Mamaladze66} and \onlinecite{Baratoff70}. For given values of $\lambda_r$, $\kappa$, $\xi_w$, $t$, and $L$, the solutions of Eq.\ (\ref{f0integral}) reveal that $\tilde j$ is a single-valued function of $f_0$, starting with the value $\tilde j = 0$ at $f_0 = 0$, initially increasing linearly with $f_0$, rising to a maximum value defined as $\tilde j_c$, then returning to zero at a larger value of $f_0$. 

When $t > 1$, the above equations reveal that $\tilde j(\phi)$ is a single-valued function of $\phi$  and has a functional dependence close to $\tilde j = \tilde j_c \sin \phi$, similar to that of a Josephson junction.  For $0 \le \phi \le \pi$, the reduced order parameter $f(\tilde j, x)$ at $x = 0$ has its maximum value $f_{00}=f(0,0)$ when $\phi = 0$, its minimum value 0 when $\phi = \pi$, and a value between these two limits at the critical current when $\tilde j = \tilde j_c$ and $\phi \approx \pi/2$.

When $T > T_{cw}$ and $L \gg \xi(T)=\xi_w/\sqrt{t-1}$, $f \ll 1$ for a large fraction of the length $L$, and one may omit the term proportional to $f^3$ on the right-hand side of Eq.\ (\ref{GL1}) to obtain the reduced order parameter $f(\tilde j,x)$.  In the absence of a current,
$f(0,x) = f_r \cosh(x/\xi)/\cosh(L/2\xi)$ near the center of the weak link, 
$f_{00} = f(0,0) = f_r/\cosh(L/2\xi) \approx 2 f_r e^{-L/2\xi}$
at the center,
and the gauge-invariant phase difference across the weak link is $\phi= 0$.  The parameter $f_r$, which is of the order of unity, would be equal to unity if the linearized GL equation were valid over the entire length $L$ of the weak link; the suppression of $f_r$ below unity occurs because the exact solution for $f(0,x)$ near $x \approx \pm L/2$ is strongly influenced by the term $(\kappa/\lambda_r)^2 f^3 $ on the right-hand side of Eq.\ (\ref{GL1}).

For nonzero current, $f_0=f(\tilde j,0)$, the reduced order parameter at the center of  the weak link,  is suppressed below $f_{00}$, and the gauge-invariant phase difference $\phi$ across the weak link obeys $\sin(\phi/2) = f_0/f_{00}.$
The reduced current is given by
\begin{equation}
\tilde j = \frac{ \lambda_r}{2 \xi}f_{00}^2\sin \phi
\approx \frac{4 f_r^2 \lambda_r}{\xi}\Big(\frac{f_0}{f_{00}}\Big)
\sqrt{1-\Big(\frac{f_0}{f_{00}}\Big)^2}e^{-L/\xi},
\label{tildejLong}
\end{equation}
such that the reduced critical current is given for $T > T_{cw}$ and any $L$ by the approximation
\begin{equation}
\tilde j_c =j_c/j_r = (\lambda_r/2 \xi)f_{00}^2\approx (2 f_r^2 \lambda_r/\xi)e^{-L/\xi}
\label{jcLong}
\end{equation}
at the maximum of $\tilde j$, where $f_0 = f_{00}/\sqrt{2}$ and $\phi = \pi/2$.  From Eq.\ (\ref{jcLong}), we may obtain the critical current as $I_c = j_c Ld = j_r \tilde j_c Ld $.  Inferring $T_{cw} = 170.9$ mK  from the experimental data in Fig.\ 1(b) for $L$ = 290 $\mu$m and assuming $\kappa = \lambda_r/\xi_w$, we obtained $\xi_w$ and $\lambda_r$ by fitting  the experimental $I_c$ data  for $L$ = 8 $\mu$m at 250 mK and 375 mK.  The dotted curves in Fig.\ 1(a) and (b) show $I_c$ calculated using $\xi_w$ = 738 nm and $\lambda_r$ = 79 nm in Eq.\ (\ref{jcLong}) and Eq.\ (\ref{f0integral}), from which $f_{00}$ was obtained.\cite{Tmax}
  
Under conditions for which Eq.\ (\ref{jcLong}) is valid, if we define the effective transition temperature $T_c(\tilde j)$ as the temperature at which the first voltage appears along the length of the TES when it carries a reduced current density $\tilde j$, we can determine $T_c(\tilde j)$ or $t_c(\tilde j)=T_c(\tilde j)/T_{cw}$ by setting $\tilde j = \tilde j_c$ in Eq.\ (\ref{jcLong}) and solving for $t_c(\tilde j)$, noting that $\xi=\xi_w/\sqrt{t-1}$.  The result is 
\begin{equation}
(T_c-T_{cw})/T_{cw}= (\xi_w^2/L^2)\ln^2(2f_r^2\lambda_r\sqrt{t_c-1}/\tilde j \xi_w).
\label{tc}
\end{equation}
Since the dependence upon $t_c$ on the right-hand side is very weak, because it appears within the argument of the logarithm, Eq.\ (\ref{tc}) predicts that the current-dependent transition temperature of the TES should scale very nearly as $T_c - T_{cw} \propto 1/L^2$ and that $T_c$ should increase as the square of the logarithm of the inverse TES current.  Similar reasoning leads to the conclusion that both $\Delta T_1$ and $\Delta T_2$ scale as $1/L^2$.
Scaling of $\Delta T_R$ can be understood using a simple model of the resistive transition based on the assumption that $R = (2x_j/L)R_N$, where, for a given reduced current density $\tilde j$, $x_j$ is the solution of $\tilde j = (\lambda_r/2 \xi)f(0,x_j)^2$.

We conclude that TESs behave as weak links.  This conclusion is based on our experimental findings that (a) the critical current at the first onset of a voltage along the length depends exponentially upon the length $L$ and the square root of the temperature $T$, (b) both the current-dependent effective transition temperature $T_c$ and the transition width scale as $1/L^2$, and (c) the TESs show clear Fraunhofer oscillations as a function of applied magnetic field, characteristic of Josephson weak links.  It follows that the strength of superconducting order is not uniform over the TES.  Our findings have implications on TES magnetic field sensitivity, which impacts required 
limits on ambient magnetic field magnitude and fluctuations in TES applications.  Proposed uses of the longitudinal proximity effect for TES applications include (1) tuning the effective $T_c$ of TES arrays by changing $L$ in mask design, which could compensate for bilayer $T_{cw}$ variability\cite{IrwinHilton05} and increase yield, and (2) making small TESs consisting of superconducting leads separated by normal metal, such as Au with $T_{cw}=0$, avoiding the use of S/N bilayers. 

Our work at Goddard was partially funded under NASA's Solar and Heliospheric Physics Supporting Research and at the Ames Laboratory  by the Department of Energy - Basic Energy Sciences under Contract No. DE-AC02-07CH11358.  We thank J.\ Beyer (PTB Berlin) and K.\ Irwin (NIST Boulder) for providing the SQUIDS used in this work.  We also thank F. Finkbeiner, R. Brekosky, and D.\ Kelly for essential roles in device fabrication, and C.\ Kilbourne, I.\ Robinson, F.\ S.\ Porter, R.\ Kelley, and M.\ Eckart for useful discussion of these results and the manuscript.


\begin{thebibliography}{99}
\bibitem{IrwinHilton05} K. D. Irwin and G. C. Hilton, in {\it Topics in Applied Physics: Cryogenic Particle Detection}, edited by C. Enss, (Springer, Berlin, 2005), p.63.
\bibitem{Bandler08} S. R. Bandler {\it et al.}, J. Low Temp. Phys. {\bf 151}, 400 (2008).
\bibitem{Bacrania09} M. K. Bacrania {\it et al.}, IEEE Trans on Nuc. Sci, {bf\ 56}, 2299 (2009).
\bibitem{Martinis00} J. M. Martinis {\it et al.}, Nucl. Instrum. Meth. A {\bf 444}, 23 (2000).
\bibitem{Luukanen03} A. Luukanen {\it et al.}, Phys. Rev. Lett. {\bf90}, 238306 (2003).
\bibitem{LindemanPerc06} M. A. Lindeman {\it et al.}, Nucl. Instrum. Meth. A {\bf559}, 715 (2006).
\bibitem{Fraser04} G. W. Fraser, Nucl. Instrum. Meth. A {\bf 523}, 234 (2004).
\bibitem{Seidel04} G. M. Seidel and I. S. Beloborodov, Nucl. Instrum. Meth. A {\bf 520},  325 (2004).  
\bibitem{SS'S} We follow the notation of Likharev\cite{Likharev79} and denote the weak link as N$'$ for $T>T_{cw}$ or S$'$ for $T<T_{cw}$.
\bibitem{Clarke69} J. Clarke, Proc. R. Soc. London, Ser. A {\bf 308}, 447 (1969).
\bibitem{Likharev79} K. K. Likharev, \rmp {\bf 51}, 101 (1979).
\bibitem{LeadsTcComment}  Measurements of TESs with Mo/Nb leads with $T_c$= 3.5 and 7.1 K were indistinguishable.
\bibitem{Chervenak04} J. A. Chervenak {\it et al.}, Nucl. Instrum. Meth. A, {\bf 520}, 460 (2004).
\bibitem{Barone82} A. Barone and G. Paterno, {\it Physics and Applications of the Josephson Effect}, (Wiley, New York, 1982).
\bibitem{Miller84} S. L. Miller and D. K. Finnemore, \prb {\bf 30}, 2548 (1984).
\bibitem{Dobrosavljevic93} L. Dobrosavljevi{\'c} and Z. Radovi{\'c}, Supercond. Sci. Technol. {\bf 6}, 537 (1993).
\bibitem{deGennes} P. G. de Gennes, {\it Superconductivity of Metals and Alloys} (Benjamin, New York, 1966), p. 177.
\bibitem{StJames69} D. Saint-James, E. J. Thomas, and G. Sarma, {\it Type II Superconductivity} (Pergamon, Oxford, 1969).
\bibitem{Pearl64} J. Pearl, \apl {\bf 5}, 65 (1964).
\bibitem{Likharev75b} K. K. Likharev and L. A. Yakobson, Sov. Phys. Tech. Phys. {\bf 20}, 950 (1975).
\bibitem{Mamaladze66} Yu. G. Mamaladze and O. D. Cheishvili, Sov. Phys. JETP {\bf 23} 112 (1966).
\bibitem{Baratoff70} A. Baratoff, J. A. Blackburn, and B. B. Schwartz, \prl {\bf 25}, 1096 (1970); errata \prl {\bf 25}, 1738 (1970).
\bibitem{Tmax} Since $\xi$ diverges at $T_{cw}$, the approximate expression for $\tilde j_c$ in  Eq.\ (\ref{jcLong}) has a local maximum at $T_{max}$ very close to $T_{cw}$.  In Fig.\ 1(a) and (b) we show calculated values of $I_c$ only for $T\ge T_{max}$.

\end{thebibliography}
\end{document}